\documentclass[12pt]{spieman}  %
\usepackage{amsmath,amsfonts,amssymb}
\usepackage{graphicx}
\usepackage{setspace}
\usepackage{tocloft}
\usepackage{lineno}
\usepackage{booktabs}  %
\usepackage{array}
\usepackage{pdfpages}     %
\newcommand{\manuscriptspacing}{1}

\title{Large language model agents for dispersion engineering of visible achromatic metalenses}

\author[a,*]{Alireza Khalilian}
\author[b]{Saeid Khalilian}
\author[a]{Hafiz Malik}
\author[c]{Yasha Yi}
\affil[a]{University of Michigan--Dearborn, Department of Electrical and Computer Engineering, Dearborn, Michigan 48128, United States}
\affil[b]{University of Passau, Faculty of Computer Science and Mathematics, Passau 94032, Germany}
\affil[c]{University of Wisconsin--Milwaukee, Department of Electrical Engineering and Connected Systems Institute, College of Engineering \& Applied Science, Milwaukee, Wisconsin 53211, United States}

\cftpagenumbersoff{figure}
\cftpagenumbersoff{table}
\begin{document}
\maketitle

\begin{abstract}
The diameter of a visible achromatic metalens is limited by the group-delay span of its meta-atom library. Here, a large language model (LLM) agent that designs TiO$_2$ meta-atoms from the simulated responses of its earlier proposals widens this span and enlarges the lens. In a prespecified benchmark with two design spaces, the agent with physics guidance widens the certified span more than the best of four non-LLM methods in all ten runs with the same number of proposals. It also leads in nine of ten at equal solver time, and removing the written physics guidance, while keeping the simulation feedback, makes no detectable difference to the span. In an exploratory search, cells designed by the agent set both ends of a library spanning 3.46\,fs, against 2.83\,fs for the standard library. This raises the nominal achromatic diameter at numerical aperture 0.2 from 16.7 to 20.5\,\textmu m and the aperture area by 50\%. In full-wave simulations, the complete lenses keep band-averaged efficiencies of 54.1--56.5\% of incident power and focal shifts under a fifth of the depth of focus across 470--670\,nm.
\end{abstract}


{\noindent \footnotesize\textbf{*}Alireza Khalilian, \linkable{akhalili@umich.edu} }

\begin{spacing}{\manuscriptspacing}   %

\section{Introduction}
\label{sec:intro}

Metasurfaces, single layers of subwavelength scatterers that shape the phase, amplitude and polarization of light, promise imaging optics that are thinner and easier to integrate than refractive ones\cite{Yu2014,Kildishev2013,Lin2014}. High-index dielectric metasurfaces combine full phase control with high transmission\cite{Arbabi2015}, and titanium dioxide (TiO$_2$) grown by atomic layer deposition has brought diffraction-limited metalenses to visible wavelengths\cite{Khorasaninejad2016,Devlin2016}. Polarization-based designs now reach high numerical aperture (NA) and efficiency in silicon-rich nitride\cite{Khalilian2025}, and metalens arrays already serve compact high-NA microscopy\cite{Khalilian2026a} and polarization-resolved imaging of plant tissue\cite{Khalilian2026b}. These systems, however, operate at a single design wavelength, because each meta-atom imprints its phase through a waveguiding or resonant response that changes with frequency. A simple metalens therefore behaves like a diffractive element and brings red and blue light to focus at different depths\cite{Arbabi2017,Aieta2015,Khorasaninejad2017}.

Dispersion engineering removes this chromatic focal shift by controlling the phase that each meta-atom imparts together with its first and second frequency derivatives, the group delay and the group-delay dispersion (GDD)\cite{Chen2018,Wang2018,Shrestha2018,Wang2017}. Light from the edge of a lens travels farther to the focus than light from its center, so an achromatic lens must delay light more at its center, by an amount that grows with the diameter. In geometric-phase (Pancharatnam--Berry) metasurfaces, rotating an anisotropic meta-atom by an angle $\theta$ shifts the phase of the light converted to the opposite circular polarization by $2\theta$ at every frequency\cite{Bomzon2002}. Rotation therefore sets the phase at the design wavelength without changing the group delay, which the cross section alone determines\cite{Chen2018}. The range of delays that a library of cross sections provides then limits the lens diameter, as delay--bandwidth and Fresnel-number bounds make quantitative\cite{Presutti2020,Engelberg2021}. Visible achromatic metalenses made of TiO$_2$ pillars a few hundred nanometers tall therefore remain a few tens of micrometers wide\cite{Chen2018,Chen2019,Hou2025}.

Such libraries are usually built by sweeping a few intuitive families, such as rectangular and elliptical pillars, single and coupled fins, rings and slotted fins\cite{Chen2018,Wang2018,Chen2019}. At the two ends of the delay range, where the aperture is decided, these families yield cells that are dim or whose phase does not vary smoothly with frequency. Inverse-design methods search richer geometries\cite{Molesky2018}, either by following gradients computed through a differentiable solver, as in topology optimization and adjoint methods\cite{Chung2020,Li2022}, or with neural networks trained on large sets of simulated structures\cite{Ma2021,Jiang2021}. They usually optimize one device or train one model rather than build an interpretable, fabrication-ready library of cells, and they are seldom compared with a simple baseline given the same number of simulations.

Large language models (LLMs) coupled to software tools offer a different route, and such agents have planned chemical syntheses and discovered new algorithms by iterating on experimental or computational feedback\cite{Boiko2023,Bran2024,RomeraParedes2024}. In metasurface design, LLMs have generated meta-atoms from natural language\cite{ZhangH2025}, carried out free-form, multiobjective and multiwavelength design in the MetaChat framework\cite{Lupoiu2025} and driven differentiable solvers\cite{Huang2025}. They have also evolved their own design workflows\cite{Huang2026}, acted as general meta-optics solvers\cite{ZhangH2026} and designed multilayer and metasurface devices\cite{Kim2025}, and transformers have served as fast surrogates in the inverse design of achromatic metasurfaces\cite{Bian2025}. LLM orchestration and physics-based evaluation are therefore not new, but a recent review finds no LLM study of achromatic or dispersion-engineered meta-atoms and calls for quantitative baselines with reported denominators\cite{ZhangReview2026}. Simple search and established black-box optimizers, such as Bayesian optimization\cite{Jones1998} and the covariance matrix adaptation evolution strategy (CMA-ES)\cite{Hansen2001}, provide such baselines. To our knowledge, however, no controlled, repeated comparison has tested whether LLM proposals widen a library more than these methods with the same number of proposals. A single design campaign cannot settle this question, because its outcome depends on the run, and an analysis chosen after the data are seen can favor any method\cite{Nosek2018}.

Here, we test an LLM agent on the dispersion engineering of visible achromatic metalenses made of 600\,nm tall TiO$_2$ pillars. We call the agent physics-grounded because it proposes new cells after studying the simulated responses of its earlier proposals. Each proposal is certified only if it passes a fixed sequence of tests by rigorous coupled-wave analysis (RCWA). An exploratory design campaign, in which the stream of LLM proposals (the LLM lane) was matched by lanes of control proposals made without a language model, supplied the primary hypothesis of a prespecified benchmark. In the benchmark, two LLM lanes, with and without physics guidance, competed with four non-LLM lanes (random mutation, a surrogate predictor, Bayesian optimization and CMA-ES) in a parametric and a free-form design space. Both LLM lanes are physics-grounded in this sense and receive the same simulation results. The physics guidance is an addition given to only one of them and consists of a written description of the task physics and its simulated trends, testable hypotheses and a notebook of their outcomes. Each lane ran in five seeds, repeated runs that start from different random subsets of a standard library, and one lane in one seed is called a lane-run. An exploratory search with the LLM lanes alone then extended the library, and four metalenses at NA~0.2 were evaluated in full-lens three-dimensional (3D) finite-difference time-domain (FDTD) simulations. In both design spaces, the LLM lane with physics guidance widened the certified delay span more than the best non-LLM lane in every seed, with the same number of proposals. It also led in nine of these ten comparisons at equal solver time, and adding the physics guidance to the simulation feedback made no detectable difference to the span. In the exploratory search, cells designed by the agent set both ends of the widest library, which enlarges the nominal achromatic diameter at $\mathrm{NA}=0.2$ from 16.7\,\textmu m for the re-certified standard library to 20.5\,\textmu m, or 19.9\,\textmu m with the end-cell delays from a full-wave FDTD check. In full-wave simulations, the larger diameter costs 2.4 percentage points of band-mean efficiency.

\section{Results and Discussion}
\label{sec:results}

\subsection{The Delay Span Limits the Aperture}
\label{sec:span}

We treat the metalens as a linear, time-invariant filter that imparts the phase $\varphi(\boldsymbol{r},\omega)$ at transverse position $\boldsymbol{r}$. With fields that carry the time dependence $e^{-i\omega t}$, a delay $\tau$ adds the phase $+\omega\tau$. All frequencies of the band share a focus at the axial distance $f$ if the imparted phase plus the propagation phase to the focus, $(\omega/c)(r^2+f^2)^{1/2}$, is the same at every point of the aperture. To second order about the design frequency $\omega_0$ (Sec.~S1 in the Supplementary Material), this condition sets three requirements. First, the phase at $\omega_0$ must follow the hyperbolic profile of an ideal lens. Second, the GDD, its second derivative with respect to frequency, must be uniform across the aperture, because the propagation phase is linear in frequency and adds no GDD of its own. Third, the first derivative of the imparted phase, the group delay $\tau = \partial\varphi/\partial\omega$, must compensate the longer path from every off-axis point,
\begin{equation}
  \tau(r) = \tau_0 - \frac{\sqrt{r^2+f^2}-f}{c},
  \label{eq:tau}
\end{equation}
so that a short pulse launched across the aperture reaches the focus at the same instant from every point. A lens for one wavelength needs only equal phase at the focus, whereas an achromatic lens also needs equal arrival times.

\begin{figure}[tbp]
\begin{spacing}{1}
  \centering
  \includegraphics[width=\textwidth]{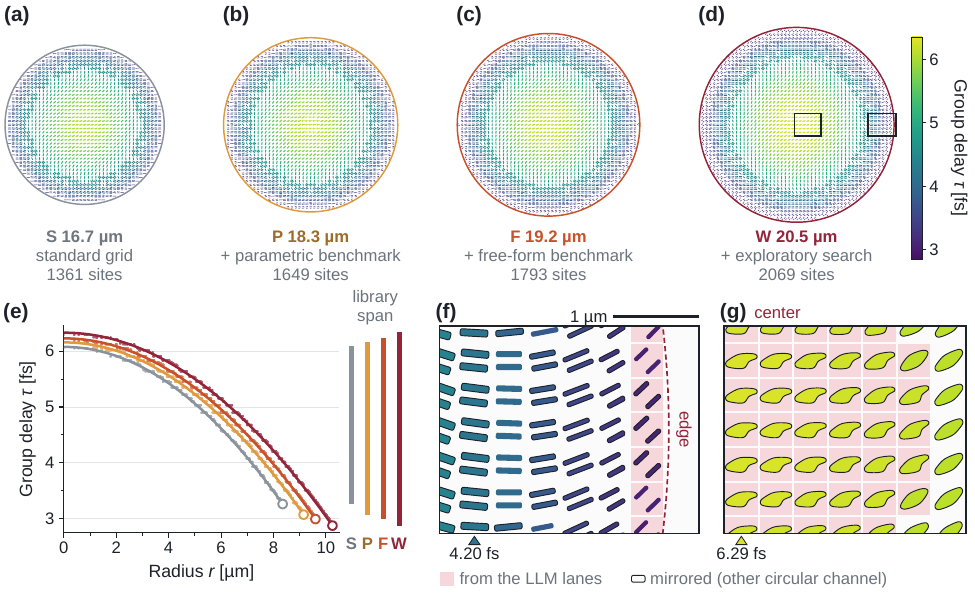}
  \caption{The size ladder ($\mathrm{NA}=0.2$, 470--670\,nm). (a)--(d)~Top views of Lens~S (16.7\,\textmu m, re-certified standard library), Lens~P (18.3\,\textmu m, with the cells of the parametric benchmark), Lens~F (19.2\,\textmu m, with the cells of the free-form benchmark) and Lens~W (20.5\,\textmu m, with the cells of the exploratory search), drawn to scale and colored by the group delay of each meta-atom. (e)~Required group delay $\tau(r)$ [lines, Eq.~\eqref{eq:tau}, with open circles at the lens edges] and delay of the cell placed at each site (dots), with bars at the right that give the contiguous span of each library. (f)~Edge and (g)~center of Lens~W [boxes in (d)], drawn at the same scale and with the lens center in the left column of (g), showing the 600\,nm tall TiO$_2$ pillars in their 400\,nm unit cells, each rotated to set the geometric phase. Shaded cells come from the LLM lanes, and outlined cells were certified in the other circular channel and are placed as mirror images. Triangles mark the two cells of Fig.~\ref{fig:cells}(a) that appear in (f) and (g), the 4.20\,fs cell ($r = 8.0$--8.1\,\textmu m) and the 6.29\,fs cell ($r < 0.9$\,\textmu m).}
  \label{fig:ladder}
\end{spacing}
\end{figure}

Because the constant $\tau_0$ is arbitrary, only the range of delays matters. With $R = D/2$ and $f = R(\mathrm{NA}^{-2}-1)^{1/2}$, the delay difference between the center and the edge of the lens is $\Delta\tau = {[(R^2+f^2)^{1/2}-f]/c}$. The delay span of a library is the range of delays that its certified cells cover without gaps larger than 0.25\,fs. A library can therefore supply every required delay only up to the diameter at which $\Delta\tau$ equals its span, and inverting the relation gives this largest nominal achromatic diameter,
\begin{equation}
  D_{\max} = \frac{2c\,\Delta\tau}{\mathrm{NA}^{-1}-\sqrt{\mathrm{NA}^{-2}-1}}
  \;\simeq\; \frac{4c\,\Delta\tau}{\mathrm{NA}} \quad (\mathrm{NA}\ll 1).
  \label{eq:dmax}
\end{equation}
At $\mathrm{NA}=0.2$, every femtosecond of span is worth 5.9\,\textmu m of diameter (diameters truncated to 0.1\,\textmu m). The re-certified standard library of Sec.~\ref{sec:campaign} spans 2.829\,fs and allows 16.7\,\textmu m, and the campaign's expanded library (3.078\,fs) allows 18.2\,\textmu m. With the certified cells of the parametric or the free-form benchmark, the standard library spans 3.099 or 3.235\,fs and allows 18.3 or 19.2\,\textmu m. The final library of the exploratory search spans 3.463\,fs and allows 20.5\,\textmu m. All but the campaign's library define a size ladder of four lenses, each as wide as its library allows (Fig.~\ref{fig:ladder}) and named S (standard), P (parametric), F (free-form) or W (widest) after its library.
The diameter $D_{\max}$ is a criterion on the span rather than on optical performance. Assigned cells may depart from the required delay, their GDD is not controlled, and a lens can exceed $D_{\max}$ with gradual degradation. We therefore call it the nominal achromatic diameter and judge performance separately by the focal spread, the Strehl ratio and the encircled energy at a common image plane.

Because a library offers only discrete delays, our lens designs assign to each site a cell within $\delta\tau = 0.04$\,fs of the required delay wherever the library allows (Sec.~\ref{sec:x-lens}). The resulting phase error, $\delta\varphi = \delta\tau\,\Delta\omega$, grows with the frequency offset $\Delta\omega = \omega-\omega_0$ from the design frequency ($\lambda_0 = 570$\,nm) but stays below 0.03\,rad over the band (Sec.~S1 in the Supplementary Material). The diffraction-limited full width at half maximum (FWHM) of the focal spot, $0.514\,\lambda/\mathrm{NA}$, is 1.2--1.7\,\textmu m here. The depth of focus, the axial range of order $\lambda/\mathrm{NA}^2$ over which the spot stays close to its smallest size, is 12--17\,\textmu m. A lens without dispersion control, which focuses at $f(\lambda) \simeq f_0\lambda_0/\lambda$, would move its focus by about a third of the focal length across the band, and these three scales serve to judge the results.

Each unit cell is a TiO$_2$ pillar of height $h = 600$\,nm on silica, on a square lattice of period $p = 400$\,nm, illuminated from the substrate. Its response is the power-normalized Jones matrix of the forward zero order, $\mathbf{J}_0(\omega)$. Rotating the cell counterclockwise by $\theta$, $\mathbf{J}(\theta) = \mathbf{R}(\theta)\,\mathbf{J}_0\,\mathbf{R}(-\theta)$, leaves the co-circular elements unchanged and multiplies the cross-circular (converted) elements, in the basis $(\hat{\boldsymbol{x}} \pm i\hat{\boldsymbol{y}})/\sqrt2$, by a factor that does not depend on frequency\cite{Bomzon2002},
\begin{equation}
  J_{\pm\mp}(\theta,\omega) = J_{\pm\mp}(0,\omega)\,e^{\mp 2i\theta},
  \qquad \tau = \frac{\partial \arg J_{\pm\mp}}{\partial\omega}, \quad \frac{\partial\tau}{\partial\theta} = 0.
  \label{eq:pb}
\end{equation}
Rotation therefore sets the phase of a site at $\omega_0$, whereas the cross section alone fixes its delay and dispersion. A single library of a few hundred shapes can thus serve every radius of the lens. Mirroring a cell ($y\to-y$) exchanges its two converted elements exactly, so a cell certified in either circular channel can serve a lens of either handedness.

The efficiency of a cell is its conversion efficiency, $\eta_{\mathrm{conv}}(\omega) = |J_{\pm\mp}|^2$, and band means such as $\bar\eta$ are arithmetic means over wavelengths evenly spaced across the band (21 samples for cells and 11 for lenses). For a birefringent cell without cross-coupling whose two eigenpolarizations are transmitted equally,
\begin{equation}
  \eta_{\mathrm{conv}}(\lambda) \simeq T(\lambda)\,\sin^2\!\left(\frac{\pi h\,\Delta n_{\mathrm{eff}}(\lambda)}{\lambda}\right),
  \label{eq:halfwave}
\end{equation}
where $T$ is the transmission and $\Delta n_{\mathrm{eff}}$ the form birefringence of the pillar, and the phase acquired at its end faces is neglected. This half-wave-plate condition anticipates the central difficulty at the short-delay end of a library, near the shortest delay of its contiguous span. There, shapes that hold little TiO$_2$ have small birefringence and convert poorly.

\begin{figure}[tbp]
\begin{spacing}{1}
  \let\footnotesize\scriptsize %
  \centering
  \includegraphics[width=0.92\textwidth]{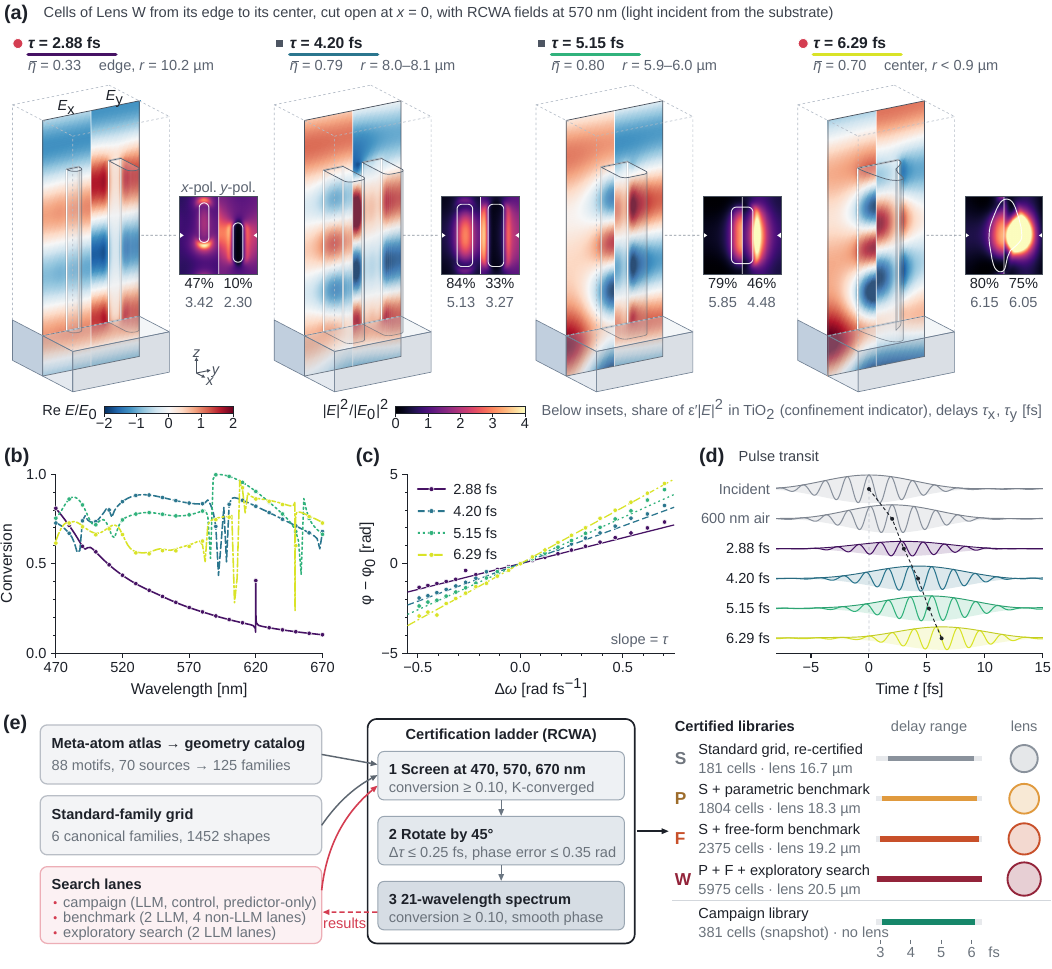}
  \caption{Certified meta-atoms and the certification ladder. (a)~Four certified cells of Lens~W, from the edge of the lens (left, $r = 10.2$\,\textmu m) to its center (right), with the field at 570\,nm for light incident from the substrate, computed with the RCWA solver of the certification (625 harmonics, Sec.~S3.6 in the Supplementary Material). Each 400\,nm unit cell (dashed box) is cut open along the plane $x = 0$, with $x$ along the long axis of the cell. The left half of the cut shows $\mathrm{Re}\,E_x$ for $x$-polarized input and the right half $\mathrm{Re}\,E_y$ for $y$-polarized input, both for unit incident amplitude $E_0$. White lines outline the TiO$_2$ in the cut, a dashed white line marks the substrate surface, and the substrate and pillar in front of the cut are translucent. Insets show $|\boldsymbol{E}|^2$ in the horizontal mid-plane of the pillar ($z = 300$\,nm) for $x$- and $y$-polarized input, seen from above with $x$, the long axis of the cell, pointing down and $y$ to the right. Below the insets are the share of $\varepsilon'|\boldsymbol{E}|^2$ that lies inside TiO$_2$ (a confinement indicator) and the delays $\tau_x$ and $\tau_y$ (fs). Labels give the certified delay $\tau$, the band-mean conversion $\bar\eta$, the origin (red circle for the exploratory search, dark gray square for the standard grid) and the radii at which Lens~W uses each cell. The colored line of each label identifies that cell in panels (b)--(d). (b)~Conversion spectra, certified at 21 wavelengths (dots) and computed with the same solver at 2\,nm wavelength steps, with finer steps of 0.05--0.1\,nm near the resonances at 620 and 650\,nm (lines). (c)~Phase of the converted element versus the frequency offset $\Delta\omega = \omega-\omega_0$ from the design frequency (dots), with lines of slope $\tau$. (d)~Transit of a pulse centered at 570\,nm (Gaussian spectral window of rms width 0.30\,rad\,fs$^{-1}$ within 470--670\,nm), computed from the spectra of (b), for the incident pulse, the pulse after 600\,nm of air and the pulse after each cell (dots, envelope peaks). The real field (lines) and envelope (shading) are normalized to the incident peak, and time is measured from the arrival of the incident pulse at the base of the pillar.
  (e)~Certification workflow, in which candidates from the literature atlas and geometry catalog, the standard-family grid and the search lanes go through the same three RCWA tests (Sec.~\ref{sec:x-ladder}) and each search lane receives the results of its own candidates. At right, the certified libraries with their contiguous delay ranges (gaps of at most 0.25\,fs) and the lens that each defines at $\mathrm{NA}=0.2$, drawn to scale (Fig.~\ref{fig:ladder}).}
  \label{fig:cells}
\end{spacing}
\end{figure}

The group delay has an equally simple origin, because a pillar behaves as a short section of waveguide\cite{Chen2018}, with delay
\begin{equation}
  \tau \simeq \frac{h}{c}\,n_{\mathrm{g}}, \qquad
  n_{\mathrm{g}} = n_{\mathrm{eff}} - \lambda\,\frac{\mathrm{d}n_{\mathrm{eff}}}{\mathrm{d}\lambda},
  \label{eq:ng}
\end{equation}
where $n_{\mathrm{eff}}$ and $n_{\mathrm{g}}$ are the effective and group indices of its guided modes averaged over the two eigenpolarizations. The delay span of a library is therefore set by the range of modal group indices that its shapes reach. With $h/c = 2.0$\,fs, the certified cells of the campaign span $n_{\mathrm{g}} = 1.52$--3.06, and those of the exploratory search span 1.43--3.17. Wide, filled pillars give the longest delays, whereas the shortest delays need modes that lie largely in air yet still convert polarization. Full-wave fields computed inside four cells of Lens~W show both regimes [Fig.~\ref{fig:cells}(a--d)].
The 2.88\,fs cell at the edge of the lens consists of two staggered LLM-proposed bars 51\,nm wide. Only 10\% of $\varepsilon'|\boldsymbol{E}|^2$, a permittivity-weighted confinement indicator, in its pillar layer lies inside TiO$_2$ for $y$-polarized input (47\% for $x$), against 75--80\% for both polarizations in the 6.29\,fs cell at the center. A pulse centered at 570\,nm leaves these two cells 1.0 and 4.3\,fs after a pulse that crosses 600\,nm of air [Fig.~\ref{fig:cells}(d)], which takes $h/c = 2.0$\,fs. Their transit times of about 3.0 and 6.3\,fs are therefore close to the certified delays of 2.88 and 6.29\,fs. The bulk group index of atomic-layer-deposited TiO$_2$ at 570\,nm, computed as $n-\lambda\,\mathrm{d}n/\mathrm{d}\lambda$ from its measured refractive index\cite{Zhukovsky2015}, is $n_{\mathrm{g}} = 2.79$. Taking this value for the upper end and unity for the lower one gives
\begin{equation}
  \Delta\tau_{\max} \approx \frac{h}{c}\left(n_{\mathrm{g}}^{\mathrm{TiO_2}}-1\right) \approx 3.6\,\mathrm{fs}
  \label{eq:ceiling}
\end{equation}
for a 600\,nm layer. This waveguide estimate is not a strict bound, because waveguide dispersion, end-face reflections and lattice effects can add delay. Resonant cells can also exceed it locally, often at the cost of the smooth dispersion that certification requires. For $\mathrm{NA}\ll1$, Eq.~\eqref{eq:dmax} and the Fresnel number $N_{\mathrm{F}} = R^2/(\lambda_0 f)$ give $N_{\mathrm{F}} \simeq 2c\,\Delta\tau/\lambda_0$, so Eq.~\eqref{eq:ceiling} corresponds to $N_{\mathrm{F},\max} \approx 3.8$, close to the 3.7 of Lens~W. General delay--bandwidth bounds\cite{Presutti2020}, expressed as a maximum Fresnel number\cite{Engelberg2021}, allow a far larger $N_{\mathrm{F}} \approx 8.4$ for 600\,nm tall TiO$_2$ structures over this band. The delay span of a library thus sets the largest aperture, whereas the conversion and phase accuracy of its cells set how much light reaches the focus (Sec.~S1.5 in the Supplementary Material).

\subsection{From a Literature Atlas to Certified Libraries}
\label{sec:campaign}

The search began from a literature-informed meta-atom atlas of 88 motif cards drawn from 70 sources (Sec.~S2.1 in the Supplementary Material). The atlas spans conventional achromatic cells\cite{Chen2018,Wang2018,Shrestha2018,Chang2024}, constructive geometry\cite{Barr1981,Gielis2003,Osher1988}, fabrication constraints\cite{Piggott2017} and resonator motifs from metamaterial physics\cite{Pendry1999,Miroshnichenko2015}, and is implemented as a geometry catalog of 125 parametric families. During its development, the agent evolved through numbered versions, V1 to V22.7 (Sec.~\ref{sec:x-ai}). Every candidate, from whatever source, must pass the same certification ladder of three RCWA tests of increasing cost [Fig.~\ref{fig:cells}(e) and Sec.~\ref{sec:x-ladder}]. A three-wavelength screen requires the cell to be bright, converting at least 10\% of the incident light into the opposite circular polarization at 470, 570 and 670\,nm, and converged in the number $K$ of Fourier harmonics. A rotation test then checks that delay and phase follow Eq.~\eqref{eq:pb} at $45^\circ$, midway between $0^\circ$ and $90^\circ$, the two orientations at which the square lattice makes Eq.~\eqref{eq:pb} exact. A 21-wavelength spectrum finally checks the conversion and the smoothness of the phase across the band. Of 50{,}274 three-wavelength screens, 44{,}814 covered 12{,}184 shapes from 225 atlas-derived families. Bright shapes were common (3638), but only 482 of them were also $K$-converged. Only 34 atlas shapes qualified under the criteria in use before version V17 [3.49--4.87\,fs, Fig.~S1(a) and Sec.~S2.3 in the Supplementary Material], so convergence and smooth dispersion, rather than brightness, were the bottleneck. A dense sweep of the six canonical families (1452 shapes) gave the campaign a standard library of 118 certified cells (3.36--6.05\,fs, a span of 2.69\,fs and 15.97\,\textmu m at $\mathrm{NA}=0.2$). Re-certified for the benchmark with every shape screened (Sec.~\ref{sec:x-bench}), the same grid yields 181 certified cells from 3.255 to 6.083\,fs, a span of 2.829\,fs and a nominal diameter of 16.7\,\textmu m. This re-certified standard library is the baseline for every comparison of span and for Lens~S. No atlas shape outside these families was certified, so we regard the atlas as a sample of the design space and a released resource rather than as evidence that the literature contains the missing cells.

In the exploratory campaign, an in-loop LLM inventor proposed cells beyond either end of the certified delay range or brighter than any certified cell of similar delay. Every proposal was matched by a random-mutation control, a cell made without a language model by randomly perturbing one of the ten best existing cells for the same target. In the final rounds, each proposal was also matched by a predictor-only candidate, the best of about 150 random mutations of the best certified cells as ranked by a surrogate predictor. The predictor is a $k$-nearest-neighbor model trained on all earlier screens (Sec.~\ref{sec:x-agent}), and these two kinds of proposals form the control and predictor-only lanes. Over the free-form rounds, 108 of 628 LLM proposals were certified (17.2\%), compared with 76 of 628 controls (12.1\%, two-sided two-proportion $z$-test, $P = 0.011$) and 44 of 514 predictor-only candidates (8.6\%). A cell is useful if it extends the delay range, is brighter than earlier cells of similar delay or is used in a lens design. The LLM lane produced 36 useful cells, against 10 for the control lane ($P = 9.4\times10^{-5}$, Fig.~S2 and Table~S2 in the Supplementary Material). Its advantage in useful cells remains significant under every paired and clustered analysis ($P \le 0.024$), whereas its advantage in certification rate does not when whole control lineages are resampled (clustered McNemar $P = 0.095$, Table~S3 in the Supplementary Material). Against the predictor-only lane, the LLM's proposals were certified about twice as often but did not differ significantly in useful cells (6.0\% versus 3.5\%, $P = 0.057$). Predictor-only cells also fill more sites of the agent's largest scalar lens design (736 against 464 of 1617), which makes a data-driven predictor a strong baseline for supplying the cells that lenses use. These exploratory comparisons come from one campaign with one inventor model and match the lanes per proposal rather than per unit of computation. Measured in solver time, the LLM lane used 5.0 worker-hours (hours of one two-thread solver process) per useful cell, against 15.4 for the control lane and 8.2 for the predictor-only lane (Sec.~S3 and Table~S4 in the Supplementary Material).

Most of the campaign's gain came from the short-delay end (Fig.~S1 in the Supplementary Material), where Eqs.~\eqref{eq:halfwave} and \eqref{eq:ng} predict the greatest difficulty. A short delay demands little TiO$_2$, yet the cell must keep enough birefringence to convert light. Most of this extension, down to 3.06\,fs, came from LLM-proposed pairs of 50\,nm wide fins separated by gaps of 80--100\,nm (mostly 100\,nm), whereas the standard-family grid used gaps of 60 and 80\,nm. These LLM-proposed pairs have group indices of 1.53--1.59, consistent with a mode largely in air, as the fields inside a related two-bar cell at the edge of Lens~W show [Fig.~\ref{fig:cells}(a)]. The long-delay end moved from 6.05 to 6.12\,fs through free-form cells of the non-LLM lanes. The expanded library spans 3.04--6.12\,fs (3.08\,fs), about 86\% of the waveguide estimate of Eq.~\eqref{eq:ceiling}, for a nominal achromatic diameter of 18.2\,\textmu m [Fig.~S1(b) in the Supplementary Material]. The short-delay cells, however, belong to a conventional family whose reference grid was incomplete. A targeted search of fin pairs under the same gates and fabrication rules (248 distinct structures) reaches 2.989\,fs, against 3.042\,fs for the LLM's cell, at a similar cost (Fig.~S3 in the Supplementary Material). We therefore report the LLM's role as the provenance of these designs rather than as evidence of an advantage over a properly extended parametric search. Because the baselines searched where the LLM chose to search and mutated a shared pool enriched with certified LLM cells, the campaign could not isolate the method. Its main observation, a wider certified delay span, therefore became the primary hypothesis of a prespecified benchmark.

\subsection{Prespecified Benchmark of LLM and Non-LLM Lanes}
\label{sec:bench}

\begin{figure}[tbp]
\begin{spacing}{1}
  \centering
  \includegraphics[width=\textwidth]{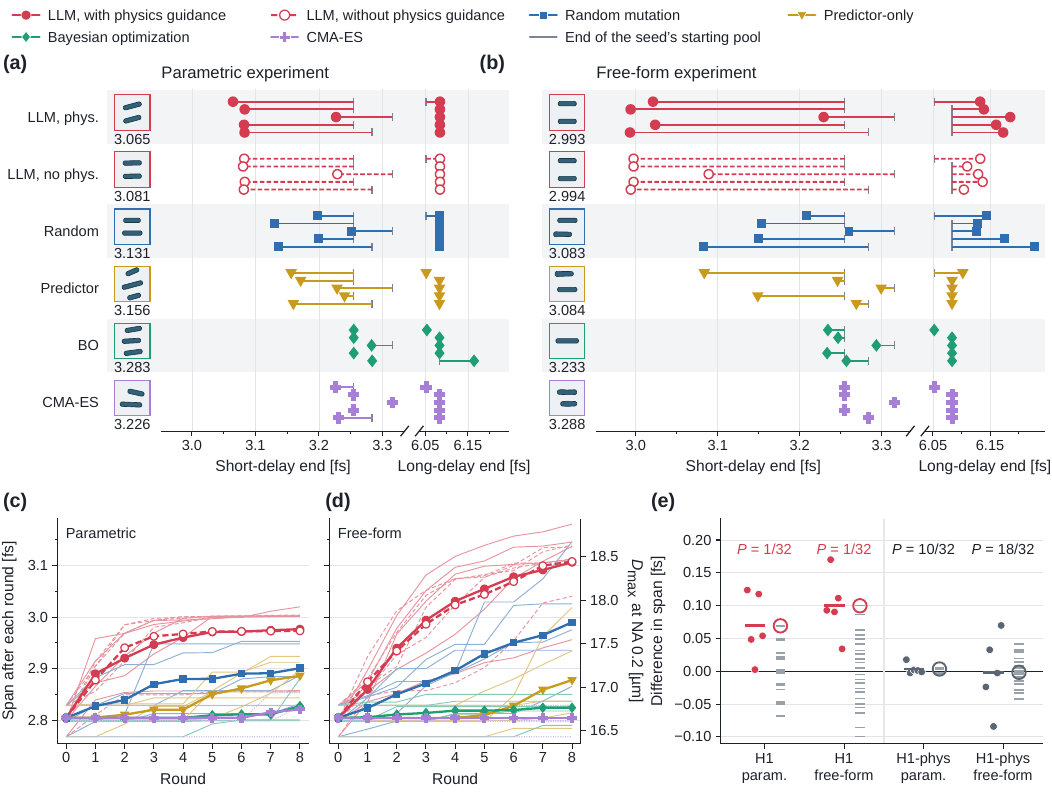}
  \caption{Prespecified benchmark of two LLM lanes against four non-LLM lanes, with five seeds (repeated runs from different random subsets of the standard library) and eight rounds of 20 proposals per lane-run (one lane in one seed). (a)~Parametric and (b)~free-form experiment, in which two horizontal segments for every lane-run show how far it moved each end of the certified delay range, the shortest delay (left) and the longest delay (right, broken axis), with seeds 1--5 from top to bottom within each lane. Each segment runs from the end of its seed's starting pool (gray tick) to the end reached after eight rounds (marker). Glyphs show the cross section of the shortest certified cell of each lane over the five seeds in its 400\,nm unit cell, labeled with its delay in fs (longer than the end of the starting pool for a lane that did not extend the short end). (c),~(d)~Contiguous certified span after each round for every lane-run (thin lines, dotted for CMA-ES) and averaged over the five seeds (thick lines with markers), with the corresponding nominal achromatic diameter at $\mathrm{NA}=0.2$ on the right axis [Eq.~\eqref{eq:dmax}]. (e)~Confirmatory tests, with the per-seed differences in final span (dots) between the LLM lane with physics guidance and either the best non-LLM lane of the seed (H1) or the LLM lane without physics guidance (H1-phys), and their mean (bar). Gray ticks show the 32 means obtained by flipping the signs of the five differences in every possible way, which define the exact one-sided test. The ring marks the observed mean, and $P$ is the fraction of the 32 means at or above it. Row labels LLM, phys. and LLM, no phys. denote the two physics-grounded LLM lanes with and without the added physics guidance, Predictor the predictor-only lane and BO Bayesian optimization.}
  \label{fig:bench}
\end{spacing}
\end{figure}

The benchmark tested whether, given the same number of proposals, the LLM lane with physics guidance widens the contiguous certified delay span of the standard library more than the best of four non-LLM lanes (H1). If H1 held, it also tested whether this lane widens the span more than the same LLM without the guidance (H1-phys). Both LLM lanes are physics-grounded, as defined in the Introduction, and differ only in the added guidance (Sec.~\ref{sec:x-bench}). Each lane-run had eight rounds of 20 proposals, and both experiments ran as frozen, with no lane-run or round rerun or replaced. In the parametric experiment, the LLM lane with physics guidance ended with spans of 2.857--3.020\,fs, against 2.855--2.953\,fs for the best non-LLM lane of each seed, and led in all five seeds (Fig.~\ref{fig:bench}).
Its mean lead was 0.069\,fs (bootstrap 95\% interval 0.031--0.107\,fs, $t$ interval 0.006--0.133\,fs), with a one-sided sign-flip $P = 1/32 = 0.031$ (Table~S6 in the Supplementary Material). In the free-form experiment, its spans were 2.956--3.180\,fs against 2.866--3.146\,fs, again a lead in every seed ($P = 1/32$), with a mean lead of 0.100\,fs (bootstrap interval 0.061--0.139\,fs, $t$ interval 0.039--0.160\,fs). H1 is therefore supported in both experiments, with modest margins of 0.41 and 0.59\,\textmu m of nominal diameter at $\mathrm{NA}=0.2$ on average. In one parametric seed, whose starting pool begins at 3.316\,fs, the latest of the five, the lead is only 0.0025\,fs, less than the agreement between our two RCWA solvers. This seed, in which the predictor-only lane did almost as well (2.855 against 2.857\,fs), is effectively a tie, and the parametric result depends on it, because counting it as a loss would give $P = 2/32 = 0.063$. The best non-LLM lane was random mutation in seven of the ten seeds and the predictor-only lane in the other three. Bayesian optimization and CMA-ES, which take small steps, widened the span by at most 0.081\,fs in any seed. At equal solver time (Sec.~\ref{sec:x-bench}), the LLM lane with physics guidance also led in all five parametric seeds and in four of the five free-form seeds. Gap rules of 0.15 and 0.35\,fs kept all ten primary differences positive (Sec.~S6 in the Supplementary Material).

The LLM lanes widened the span mainly at the short-delay end [Fig.~\ref{fig:bench}(a,b)], and in the parametric experiment, the lane with physics guidance moved the short end by 0.088--0.201\,fs but the long end by at most 0.033\,fs. Both LLM lanes converged on the cells that had extended the campaign's library, pairs of 50\,nm wide fins about 200\,nm long at the largest fin gap of the parametric families, 100\,nm. In the free-form experiment, both ends moved, and the shortest cells of both LLM lanes (2.993--2.994\,fs) are two parallel bars, 50\,nm wide and about 200\,nm long. These bars, spaced evenly across the cell, form the structure that the conventional fin-pair search identified. Consistent with Eq.~\eqref{eq:halfwave}, the LLM lanes' short-end cells convert about 35\% of the light on band average, against 36--44\% for the less extended short ends of the other lanes (Sec.~S6 in the Supplementary Material).

The added physics guidance, the written description of the task physics with its hypotheses and notebook, made no detectable difference to the span in either experiment. The two LLM lanes ended within 0.02\,fs of each other in every parametric seed (H1-phys, $P = 10/32 = 0.31$) and differed in both directions by up to 0.084\,fs in the free-form experiment ($P = 18/32 = 0.56$). Their spans after each round followed nearly the same course [Fig.~\ref{fig:bench}(c--e)], and some secondary endpoints differed descriptively (Sec.~S6 in the Supplementary Material). The lane without the guidance also led the best non-LLM lane in all five parametric seeds (in one by only 0.0005\,fs) and in four of the five free-form seeds. This null result concerns only the added guidance, not the grounding in simulation that both LLM lanes share. The benchmark cannot tell how much of their lead this grounding explains, because the non-LLM lanes also learn from their own simulations and no LLM lane without simulation feedback was tested. The lead therefore belongs to the LLM lanes as implemented, which, besides the model, share an agent developed on this task during the campaign and a two-stage selection. In this selection, 60 drafts are ranked by the surrogate and compared with their nearest simulated neighbors before 20 are chosen. Bayesian optimization and CMA-ES, by contrast, were configured in two-round pilots, and the benchmark does not separate these components. The LLM lanes also used the most solver time, 10--13 nominal solver hours per lane-run against 2--10\,h for the others (Sec.~\ref{sec:x-bench}), and their proposals cost US\$2.1--2.7 per lane-run at list prices.

\subsection{Exploratory Search for a Wider Library}
\label{sec:search}

\begin{figure}[tbp]
\begin{spacing}{1}
  \centering
  \includegraphics[width=\textwidth]{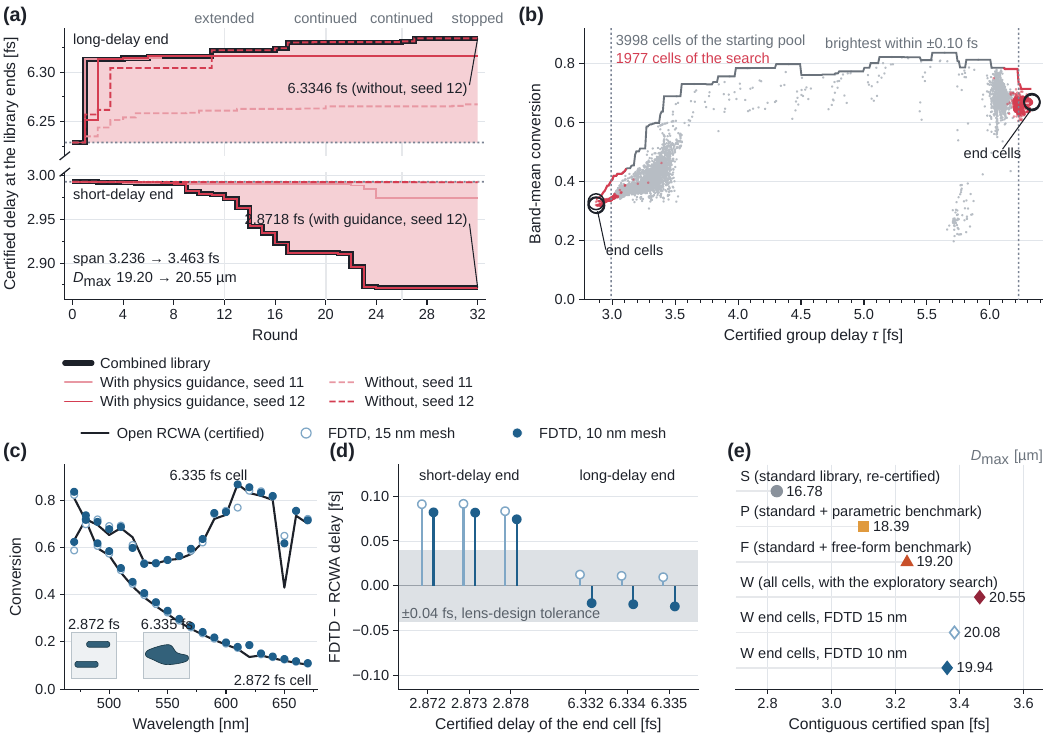}
  \caption{Exploratory search for a wider library and full-wave check of its end cells. (a)~Short- and long-delay ends of the certified library after each round, for each of the four lane-runs (its own cells together with the shared starting pool) and for the combined library (thick black line, drawn beneath the lane-runs, with labels naming the lane-run that sets each final end). Red lines show the lane-runs with physics guidance (solid) and without it (dashed), light for seed 11 and dark for seed 12, two seeds that the benchmark did not use. Dotted lines mark the ends of the starting pool, which merged every cell of the benchmark, shading marks the gain, and labels mark the decisions after rounds 12, 20, 26 and 32 (Sec.~\ref{sec:x-search}). (b)~Band-mean conversion versus certified delay for the certified cells of the starting pool (gray) and of the search (red), with the brightest cell within $\pm0.10$\,fs of each delay (lines, red where the search changed it), the six end cells (rings) and the ends of the starting pool (dotted lines). (c)~Conversion spectra of the shortest (2.872\,fs) and longest (6.335\,fs) certified cells from the open RCWA solver of the certification (lines) and from Lumerical FDTD with a 15\,nm (open symbols) and a 10\,nm (filled symbols) mesh, with their cross sections in the 400\,nm cell as insets. (d)~Difference between the FDTD and RCWA delays of the three shortest and three longest cells, with the $\pm0.04$\,fs cell-selection window of the lens designs as a gray band. (e)~Contiguous certified span and nominal achromatic diameter at $\mathrm{NA}=0.2$ (truncated to 0.01\,\textmu m) of the four libraries of the size ladder and of library W with the delays of its end cells taken from FDTD at 15 and 10\,nm mesh. The libraries are S (re-certified standard grid), P and F (with the cells of the parametric or free-form benchmark) and W (with the cells of the exploratory search).}
  \label{fig:search}
\end{spacing}
\end{figure}

Pooling every cell of the benchmark and its pilots already gave a library of 2.993--6.229\,fs (3.236\,fs, $D_{\max} = 19.20$\,\textmu m), wider than that of any single lane-run. Its short end came from an LLM lane and its long end from the random-mutation lane. From this pool, the four LLM lane-runs of the exploratory search moved both ends outward over 32 rounds, to 2.872--6.335\,fs [Fig.~\ref{fig:search}(a)].
The resulting span of 3.463\,fs, close to the waveguide estimate of 3.6\,fs [Eq.~\eqref{eq:ceiling}], gives a nominal achromatic diameter of 20.55\,\textmu m at $\mathrm{NA}=0.2$. Of the 2560 proposals, 2435 were simulated and 1977 certified, with the shortest cells from a lane-run with physics guidance and the longest from one without it. With two seeds per lane and no controls, however, this search compares neither the lanes nor the methods. Most of the new cells lie at the two ends, where the conversion falls [Fig.~\ref{fig:search}(b)]. On band average, it drops to 0.32--0.33 for the end cells at the short end, as Eq.~\eqref{eq:halfwave} anticipates for so little TiO$_2$, and to 0.67 at the long end.

The end cells are variants of two geometries, each differing by at most 1\,nm [Fig.~\ref{fig:search}(c)]. At the short end (2.872--2.878\,fs), two straight bars 50\,nm wide and about 200\,nm long, 124\,nm apart and staggered by about 100\,nm, fill 12\% of the cell. They reach delays 0.12\,fs shorter than the evenly spaced pairs of the benchmark (2.993\,fs), although which of the two changes matters was not tested. At the long end (6.332--6.335\,fs), a single smooth pillar 374\,nm long and up to 173\,nm wide, without a mirror axis, fills 26\% of the cell. Its certified converted element would require mirrored placement in our lens designs.

Full-wave FDTD confirms the extended range, with a systematic offset at the short end [Fig.~\ref{fig:search}(c--e)]. The conversion spectra agree except near 620--650\,nm, where the RCWA spectra of these cells show a one-wavelength excursion of the phase. This excursion puts their RCWA fit residuals (0.083--0.099\,rad) just below the 0.10\,rad gate, whereas the FDTD residuals are 0.005--0.046\,rad, so these cells passed certification narrowly. The FDTD delays are longer by 0.074--0.082\,fs at the short end (10\,nm mesh, or 0.083--0.091\,fs at 15\,nm) and within 0.023\,fs of the RCWA values at the long end. With the FDTD delays of its end cells, library W therefore spans 3.361\,fs at 10\,nm mesh and 3.384\,fs at 15\,nm, 97.0\% and 97.7\% of the RCWA value. The corresponding nominal diameters are 19.94 and 20.08\,\textmu m instead of 20.55\,\textmu m. The short-end offset lies outside the $\pm0.04$\,fs window of the lens designs [Fig.~\ref{fig:search}(d)]. Lens~W does not use these six cells, but its outermost cells come from the same lane-run, and its full-lens simulation, which computes the response of every cell in situ, includes any such offset.
Refitting every cell without its wavelength of largest residual shortens the span of library W to 3.400\,fs (20.18\,\textmu m) but changes those of libraries S, P and F by at most 0.025\,fs, and lenses wider than $D_{\max}$ degrade gradually (Sec.~S5 in the Supplementary Material).

\subsection{Full-Wave Performance of the Size Ladder}
\label{sec:fullwave}

\begin{figure}[tbp]
\begin{spacing}{1}
  \centering
  \includegraphics[width=\textwidth]{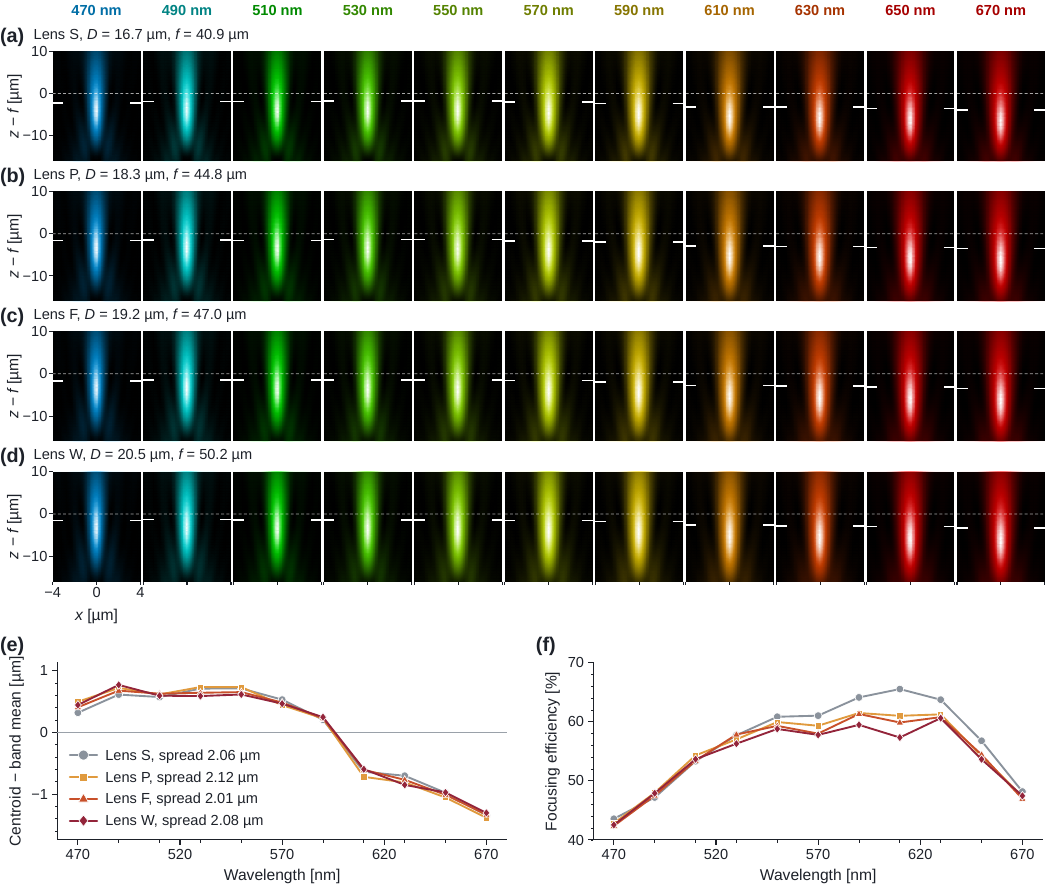}
  \caption{Full-wave focusing of (a)~Lens~S, (b)~Lens~P, (c)~Lens~F and (d)~Lens~W of the size ladder. Intensity of the converted circularly polarized field in the $x$--$z$ plane through the lens axis, propagated from the 3D FDTD near field by the angular-spectrum method, for 11 wavelengths from 470 to 670\,nm, around the design focal length $f$ ($z$ measured from the pillar tops). Each panel is normalized to its own peak intensity and drawn in the hue of its wavelength (nonlinear display), dashed lines mark the design focal plane, and short ticks mark the focal position (centroid of the main on-axis lobe above half its peak). (e)~Focal centroid relative to its band mean, with the focal spread of each lens in the legend. (f)~Focusing efficiency in the plane of the on-axis intensity maximum of each wavelength, defined as the fraction of the incident power within a radius of 1.5 times the simulated FWHM there.}
  \label{fig:fdtd}
\end{spacing}
\end{figure}

We simulated each complete lens of the size ladder (1361, 1649, 1793 and 2069 sites) in 3D FDTD under circularly polarized illumination (Fig.~\ref{fig:fdtd} and Sec.~\ref{sec:x-lens}).
On band average, they focus 54.1--56.5\% of the incident power and 61.2--64.9\% of the transmitted power into the focal spot. Relative to the incident power, the efficiency is lowest at 470\,nm (42.3--43.5\%) [Fig.~\ref{fig:fdtd}(f)]. Over the band, the focal centroid moves by 2.01--2.12\,\textmu m, 4.3--5.4\% of the mean focal distance and well inside the depth of focus of 12--17\,\textmu m [Fig.~\ref{fig:fdtd}(e)]. The mean foci lie 4.0--6.2\% short of the design focal lengths, and the diffraction focal shift of an ideal aperture at the same Fresnel numbers (3.0--3.7) accounts for 82--87\% of this shortfall (Sec.~S4 in the Supplementary Material).

Because the four lenses share the design rule, the platform and the simulation settings and differ only in their libraries, the ladder isolates what a wider span buys. From 16.7 to 20.5\,\textmu m, the diameter grows by 22.7\% and the area by 50.6\%, while the band-mean efficiency falls from 56.5 to 54.1\%. The focal spread stays at 2.01--2.12\,\textmu m, so its fraction of the focal distance falls from 5.4 to 4.3\%, and the worst-wavelength Strehl ratio at the common plane stays at 0.90--0.92. The cells from the LLM lanes sit at the edge and the center of Lenses~P, F and W [Fig.~\ref{fig:ladder}(f,g) for Lens~W], where the required delays approach the ends of the libraries, and occupy 4.4, 8.9 and 13.6\% of the sites. They are also among the dimmest cells, with mean certified conversions of 0.517--0.573 against 0.746--0.757 for the standard grid. In the lens, they keep 69--74\% of their certified conversion, against 87--89\% for the standard grid and 82--83\% for the cells of the non-LLM benchmark lanes. Because no certification gate bounds the GDD, the assigned cells carry GDD values of 0.4--2.6\,fs$^2$ (5th to 95th percentiles over the sites). The residual pupil phase of the full-wave near fields, 0.077--0.089\,rad rms on band average, nevertheless stays well below the 0.45\,rad of the Mar\'echal criterion (Sec.~S4.6 in the Supplementary Material).

\begin{figure}[tbp]
\begin{spacing}{1}
  \centering
  \includegraphics[width=\textwidth]{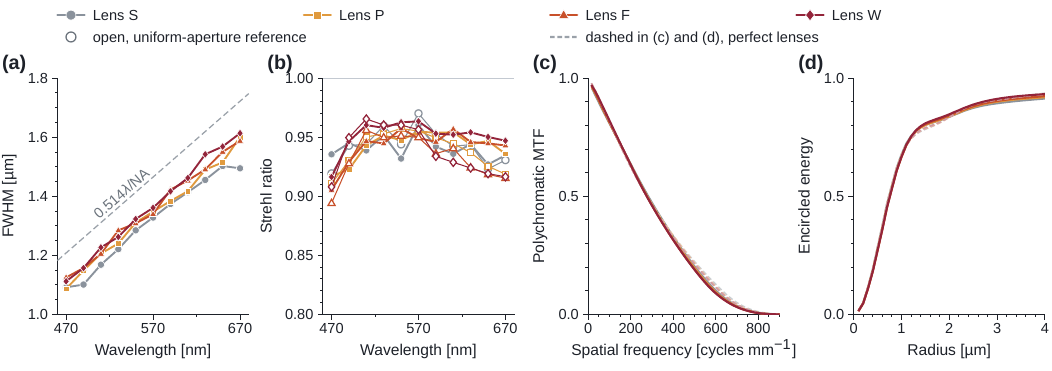}
  \caption{Full-wave imaging metrics of the size ladder, computed directly from the complex FDTD near fields without phase fitting or cell averaging. (a)~FWHM of the focal spot at the on-axis intensity maximum, with $0.514\lambda/\mathrm{NA}$ as a dashed line. (b)~Strehl ratio at the common image plane (band mean of the focal centroids), relative to a perfect-phase field with the amplitude of the converted near field inside the lens aperture widened by 0.2\,\textmu m (filled symbols) or to an aberration-free, uniformly illuminated aperture of the same converted power (open symbols, thin lines), each reference taken at its own on-axis maximum. (c)~Polychromatic modulation transfer function (MTF) at the common plane (equal energy per wavelength) with an ideal circular analyzer and (d)~polychromatic encircled energy of the converted light at the common plane, both compared with perfect lenses of the same diameters and focal lengths, each wavelength at its own focus (dashed).}
  \label{fig:direct}
\end{spacing}
\end{figure}

To assess on-axis imaging without fitting or averaging, we computed point-spread functions (PSFs) directly from the complex near fields and propagated them to the common image plane, the band mean of the focal centroids (Fig.~\ref{fig:direct}).
There, the Strehl ratio relative to a perfect-phase field of the same amplitude averages 0.94--0.95 over the band, with a minimum of 0.90--0.92 at 470\,nm (650\,nm for Lens~S). Relative to an aberration-free, uniformly illuminated aperture of the same converted power, it averages 0.93--0.94 (minimum 0.89--0.91) [Fig.~\ref{fig:direct}(b)]. With an ideal circular analyzer, the polychromatic MTF50, the spatial frequency at which the modulation transfer function (MTF) falls to 0.5, is 275--282\,cycles\,mm$^{-1}$. It is close to that of perfect lenses of the same diameters and focal lengths (274--279\,cycles\,mm$^{-1}$), and the encircled energy nearly matches theirs [Fig.~\ref{fig:direct}(c,d)]. The band-mean FWHM lies 6.7--10.4\% below $0.514\lambda/\mathrm{NA}$, partly because the foci lie closer to the lens than $f$ [Fig.~\ref{fig:direct}(a)]. Without an analyzer, the unconverted light (18--23\% of the incident power on band average) forms a broad background that lowers the MTF50 to 41--47\,cycles\,mm$^{-1}$ (Sec.~S4.3 in the Supplementary Material).

\subsection{Comparison with Simulated Achromatic Metalenses}
\label{sec:comparison}

Because the required delay scales with size and NA, Table~\ref{tab:comparison} compares simulated broadband visible achromatic metalenses by the time--bandwidth product $\Omega\,\Delta\tau$, with $\Delta\tau = {[(R^2+f^2)^{1/2}-f]/c}$ (Sec.~\ref{sec:span}) and $\Omega$ the angular-frequency width of the band.
\begin{table}[tbp]
\begin{spacing}{1}
\centering
\caption{Simulated performance of broadband visible achromatic metalenses}
\label{tab:comparison}
\renewcommand{\arraystretch}{1.15}
{\scriptsize
\setlength{\tabcolsep}{1.2pt}%
\begin{tabular*}{\textwidth}{@{\extracolsep{\fill}}l>{\raggedright\arraybackslash}p{3.0cm}ccccccccc>{\centering\arraybackslash}p{1.7cm}cc@{}}
\toprule
Ref. & Platform & $h$ & $p$ & $w_{\min}$\textsuperscript{a)} & Band & $D$ & NA & $\Delta\tau$\textsuperscript{b)} & $\Omega\,\Delta\tau$\textsuperscript{b)} & Method & $\bar{\eta}$\textsuperscript{c)} & $\Delta f/f$\textsuperscript{d)} & Pol.\textsuperscript{e)} \\
 & & [nm] & [nm] & [nm] & [nm] & [\textmu m] & & [fs] & & & [\%] & [\%] & \\
\midrule
\multicolumn{14}{@{}l}{\textit{2D or axisymmetric simulations}} \\
\citenum{Chung2020}  & TiO$_2$, inverse design & 250 & -- & 25 & 450--700 & 12.5 & 0.1/0.9 & 1.05/13.1 & 1.6/19.5 & 2D FDTD & 65/23 (m) & ${<}$DOF & TM \\
\citenum{Maher2023}  & TiO$_2$, topology opt.  & 250 & -- & n/r & 400--800 & 6 & 0.7/0.9 & 4.1/6.3 & 9.6/14.8 & 2D FEM & 62.1/47.5 (m) & n/r & L \\
\citenum{Pan2023}    & Polymer, 3 layers       & $3\times1200$ & -- & 200 & 400--800 & 20 & 0.5/0.7 & 8.9/13.6 & 21/32 & 2D FDTD & 44.2/20.0 (n/s) & n/r & U \\
\multicolumn{14}{@{}l}{\textit{Full-lens simulations}} \\
\citenum{Wang2023}   & TiO$_2$ fins (PB)       & 600 & 320 & 70 & 428--652 & 40 & 0.075 & 2.5 & 3.8 & Scalar\textsuperscript{f)} & 52 (ap) & ${<}5$ & C \\
\citenum{LiuM2021}  & TiO$_2$ fins (unrotated) & 600 & 250 & 50 & 450--800 & 10 & 0.29\textsuperscript{g)} & 2.45 & 4.5 & FDTD & 64 (i) & 0.5\textsuperscript{g)} & L \\
\citenum{Hou2025}    & TiO$_2$ fins (0$^\circ$/90$^\circ$)\textsuperscript{h)} & 600 & 400 & 50 & 450--650 & 20 & 0.164 & 2.75 & 3.5 & FDTD & 80.5 (ap) & 4.0 & I \\
\citenum{Xiao2026}  & TiO$_2$, C$_4$-symmetric pillars & 600 & 320 & 60 & 400--750 & 12.1\textsuperscript{g)} & 0.29 & 3.00 & 6.6 & FDTD & ${>}40$ (i) & n/r & I \\
\citenum{Yang2026}  & GaN, square air holes & 600 & 200 & n/r & 450--700 & 8.6 & 0.39 & 2.91 & 4.4 & 3D FDTD & 46--68\textsuperscript{i)} (m) & 8.8\textsuperscript{i)} & I \\
\citenum{Li2024}     & TiO$_2$ bilayer         & 2100 & n/r & 50/80 & 480--680 & 110 & 0.10 & 9.2 & 10.6 & FDTD & 83.6 (n/s) & 3.4 & I \\
\citenum{ZhangY2025} & Si$_3$N$_4$ pillars     & 1300 & 350 & 40 & 460--650 & 30 & 0.155 & 3.9 & 4.7 & FDTD & 91.2 (i) & ${<}1.6$ & I \\
\multicolumn{14}{@{}l}{\textit{This work (3D FDTD of the complete lens, 11 wavelengths)}} \\
Lens S & TiO$_2$ pillars (PB) & 600 & 400 & 50 & 470--670 & 16.7 & 0.20 & 2.81 & 3.37 & 3D FDTD & 56.5 (i) & 5.4 & C \\
Lens P & TiO$_2$ pillars (PB) & 600 & 400 & 50 & 470--670 & 18.3 & 0.20 & 3.08 & 3.69 & 3D FDTD & 55.1 (i) & 5.0 & C \\
Lens F & TiO$_2$ pillars (PB) & 600 & 400 & 50 & 470--670 & 19.2 & 0.20 & 3.23 & 3.87 & 3D FDTD & 54.7 (i) & 4.5 & C \\
Lens W & TiO$_2$ pillars (PB) & 600 & 400 & 50 & 470--670 & 20.5 & 0.20 & 3.45 & 4.13 & 3D FDTD & 54.1 (i) & 4.3 & C \\
\bottomrule
\end{tabular*}}

\smallskip
{\footnotesize\raggedright
Experimental values are not listed, so that all entries in the table are simulated. Because definitions of efficiency and focal deviation differ between works [footnotes c) and d)], the entries are not a ranking. Values separated by a slash are two designs of the same work or, for $w_{\min}$ in Ref.~\citenum{Li2024}, its top and bottom layers. Lens~S uses the re-certified standard library, Lenses P and F add the cells of the parametric or free-form benchmark, and Lens~W uses all certified cells, including those of the exploratory search (Sec.~\ref{sec:x-lens}). Abbreviations are $h$ = structure height, $p$ = lattice period (-- for designs without a unit-cell lattice), $w_{\min}$ = minimum feature width, n/r = not reported, n/s = not stated, DOF = depth of focus, FEM = finite-element method, opt. = optimization, Pol. = polarization and PB = Pancharatnam--Berry.\\
\textsuperscript{a)}Smallest feature width allowed by the stated fabrication rules or, where no rule is stated, smallest width of the parameter sweep (Refs.~\citenum{Wang2023,LiuM2021}). In this work, solid features are at least 50\,nm wide, voids and gaps within a cell at least 60\,nm and gaps between neighboring cells at least 20\,nm (Sec.~S2.3 in the Supplementary Material).\\
\textsuperscript{b)}Group-delay span required by the aperture, $\Delta\tau = [(R^2+f^2)^{1/2}-f]/c$, and its product with the angular-frequency width of the band, $\Omega$, computed here from the reported $D$, NA and band.\\
\textsuperscript{c)}Band-mean focusing efficiency, normalized to the incident power (i), the transmitted power (t), the power through the lens aperture (ap) or the incident power with the spot bounded by the first minimum (m). For all other entries, the spot is $3\times\mathrm{FWHM}$ ($6\times\mathrm{FWHM}$ in Ref.~\citenum{LiuM2021}), and n/s marks a normalization that is not stated.\\
\textsuperscript{d)}Focal-length deviation over the band as reported (maximum $|\Delta f|/f$, except $(f_{\max}-f_{\min})/\bar f$ in Ref.~\citenum{LiuM2021}), and for this work $(f_{\max}-f_{\min})/\bar f$ of the main-lobe focal centroid.\\
\textsuperscript{e)}C = circular (geometric phase), I = polarization-insensitive, L = single linear polarization, TM = transverse magnetic (2D), U = unpolarized.\\
\textsuperscript{f)}Fresnel--Kirchhoff integration over the whole lens of the simulated unit-cell responses.\\
\textsuperscript{g)}Computed here from the reported diameter (10\,\textmu m), mean focal length (16.63\,\textmu m) and focal shift (0.09\,\textmu m) of Ref.~\citenum{LiuM2021}, and from the NA and focal length (20\,\textmu m) of Ref.~\citenum{Xiao2026}.\\
\textsuperscript{h)}Main design of Ref.~\citenum{Hou2025} ($D = 20$\,\textmu m), optimized at three wavelengths (450, 532 and 635\,nm), with fins at $0^\circ$ or $90^\circ$ whose co- and cross-polarized components are focused together, which makes it polarization-insensitive for circular input.\\
\textsuperscript{i)}Largest of the three lenses of Ref.~\citenum{Yang2026} (target focal length 10\,\textmu m), with the efficiency range over the band, as no band mean is reported, and the peak-to-peak focal variation relative to the target focal length. Its phase profile provides about 2.3\,fs, because its focus forms at 0.8 of the design focal length (Sec.~S4.8 in the Supplementary Material).\par}
\end{spacing}
\end{table}
Lenses P, F and W require $\Omega\,\Delta\tau = 3.69$, 3.87 and 4.13, within the range of full-lens simulations of single-layer TiO$_2$ and GaN lenses (3.5--6.6)\cite{Hou2025,Wang2023,LiuM2021,Xiao2026,Yang2026}, and Lens~S slightly less (3.37). Those single-layer lenses share our 600\,nm height and, where stated, allow minimum feature widths of 50--70\,nm, and Lenses P, F and W require wider delay spans (3.08--3.45 against 2.45--3.00\,fs) over a narrower frequency band. The design of Ref.~\citenum{Hou2025} shares our height, period and minimum feature width, and at a similar diameter (20.5 against 20\,\textmu m) Lens~W reaches an NA of 0.20 instead of 0.164, which requires a 25\% wider delay span. The efficiency of Lens~W is lower than that of Ref.~\citenum{Hou2025}, whose fins focus the co- and cross-polarized light together, while the focal deviations of the two designs are similar. Our four lenses lie below the values of 1300\,nm tall Si$_3$N$_4$ pillars (4.7)\cite{ZhangY2025} and a TiO$_2$ bilayer (10.6)\cite{Li2024}. Several works report higher efficiencies or smaller focal deviations, but definitions differ, and the table does not rank the designs.

\subsection{Optical-Design Analysis}
\label{sec:optdesign}

\begin{figure}[tbp]
\begin{spacing}{1}
  \centering
  \includegraphics[width=\textwidth]{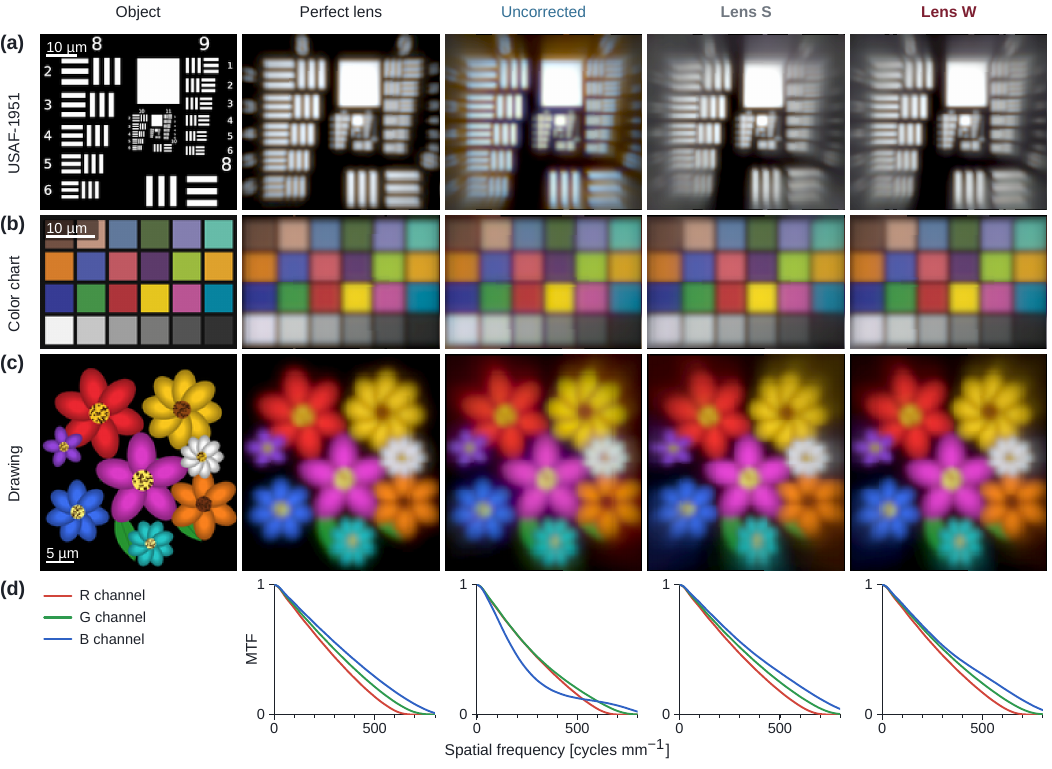}
  \caption{Zemax OpticStudio image simulation of a perfect lens, an uncorrected geometric-phase metalens and Lenses S and W. (a)~USAF 1951 chart (groups 8--11), (b)~color patches (24-patch color chart, 40\,\textmu m wide) and (c)~color drawing (original artwork, 36\,\textmu m wide), simulated at each of 11 wavelengths (Image Simulation, $9\times9$ diffraction PSF grid, pupil and image sampling $128\times128$) at each lens's common image plane. The perfect lens is an ideal paraxial surface of the same NA ($D = 20.5$\,\textmu m, $f = 50.2$\,\textmu m), and the uncorrected metalens has the diameter and focal length of Lens~W and its 570\,nm phase at every wavelength. (d)~Red-, green- and blue-channel MTFs on axis (Huygens MTFs of Fig.~S4 in the Supplementary Material, weighted by the sRGB channel sensitivities). For display, the 11 monochromatic images of each object are weighted by the source spectrum, combined into CIE~1931 XYZ values\cite{CIE2018,Wyman2013}, white-balanced to the equal-energy white of the object and encoded with the sRGB transfer function\cite{IEC1999}, identically for every column. The simulations include field-dependent aberrations but not the amplitude and efficiency variations of the FDTD fields, the angular response of the cells or the unconverted light. With these omissions and the per-wavelength scaling, which removes throughput-related color, these images do not validate physical wide-field performance or throughput fidelity. The half-widths of the charts subtend field angles of up to $40^\circ$ for Lens~S and $33$--$34^\circ$ for the other lenses (USAF chart, 64\,\textmu m wide) and $21$--$28^\circ$ (color chart and drawing), and their corners extend further, beyond the $20^\circ$ of Fig.~S4 in the Supplementary Material. The blur near their edges is the off-axis aberration of the hyperbolic phase, which the paraxial perfect lens does not have.}
  \label{fig:imaging}
\end{spacing}
\end{figure}

For field-dependent analysis and image simulation, we fitted the phase of the converted FDTD field at each wavelength with a six-term Binary~2 surface (residuals 0.019--0.112\,rad rms). We then analyzed the lenses in Zemax OpticStudio (Sec.~\ref{sec:x-zemax}) against a perfect lens and an uncorrected geometric-phase metalens (Fig.~\ref{fig:imaging}).
This phase-surface model omits the amplitude and efficiency variations of the fields and the unconverted light. In image simulations of a resolution chart, a color chart and a color drawing, Lenses S and W render the charts without the colored fringes of the uncorrected metalens. Their red-, green- and blue-channel MTF50 lie within 5\% of those of the perfect lens, whereas the green- and blue-channel MTF50 of the uncorrected lens fall 11\% and 46\% below. At $20^\circ$ off axis (Fig.~S4 and Sec.~S4.4 in the Supplementary Material), the band-mean Strehl ratio of the four lenses falls to 0.81--0.84. At 570\,nm, where the uncorrected lens is an ideal hyperbolic lens, its Strehl ratio at $20^\circ$ (0.80) is close to that of Lens~W (0.83), so these off-axis aberrations are those of the hyperbolic phase profile. Because the phase-surface model neglects the angular response of the cells, these results do not validate the designs at oblique incidence.

\subsection{Limitations and Outlook}
\label{sec:limits}

The benchmark was prespecified and frozen with hashes but not deposited in a public registry, and its free-form experiment was designed while the parametric runs were in progress. Its conclusions are limited to one inventor model, one platform, one budget of 160 proposals and one setting of each non-LLM method. The lead belongs to the LLM lanes as implemented rather than to the language model alone, and the parametric result depends on one effectively tied seed. Optimizers with longer runs, other encodings or other step sizes might close the gap to the LLM lanes. In the parametric experiment, the largest fin gap of the parametric families, 100\,nm, bounded the short end of every lane. All benchmark certification used our open RCWA solver, and the protocol's Lumerical RCWA recheck of the end cells of every lane-run was not carried out. The FDTD check of the final library found geometry-dependent offsets between the solvers (+0.074 to +0.091\,fs for thin bars, within 0.023\,fs for a wide pillar) that are as large as the margins of the benchmark. The null result for the written physics guidance shows no detectable effect on the span and says nothing about the value of physical reasoning or of the simulation feedback in general. The campaign and the exploratory search are single runs, the latter without controls.

The size ladder compares lenses at the nominal limits of their libraries rather than the largest usable lens of each. The lenses were designed in the locally periodic approximation\cite{Pestourie2018} and evaluated in full-wave simulations for one circular polarization at normal incidence. Their off-axis performance was estimated only from fitted phase surfaces that neglect the angular response of the cells (Fig.~S4 in the Supplementary Material). Fabrication tolerance was assessed only for representative campaign cells, by edge offsets of $\pm2.5$ and $\pm5$\,nm rather than with a process model (Sec.~S2 in the Supplementary Material). The narrow fin pairs at the short-delay end shift by 0.07--0.08\,fs per nanometer of edge offset, about twice the $\pm0.04$\,fs design window. The untested 50\,nm bars at the short end of the exploratory library are likely to be at least as sensitive. This study is computational, and fabricating and measuring the lenses, which this sensitivity will make demanding, is the next step. Equation~\eqref{eq:ceiling} also indicates the way forward, because the delay ceiling grows with the structure height and the group index of the material. This favors taller pillars and platforms of tunable index, such as silicon-rich nitride\cite{Khalilian2025}. Smaller periods raised the focusing efficiency of polarization-based metalenses in our earlier study\cite{Khalilian2025}, so a period below 400\,nm could raise the efficiency of the present lenses as well. A smaller period would, however, leave less room for the fin pairs that set the short-delay end of the library, so the library would have to be certified again and its delay span might narrow.

\section{Conclusion}
\label{sec:conclusion}

In summary, we have tested a physics-grounded LLM agent, which designs new meta-atoms from the simulated responses of its earlier proposals, for the dispersion engineering of visible achromatic metalenses. In a prespecified benchmark with two design spaces, the agent with physics guidance widened the certified group-delay span of a TiO$_2$ meta-atom library more than the best of four non-LLM search methods in all five seeds of each experiment with the same number of proposals. Its lead averaged 0.069 and 0.100\,fs (0.41 and 0.59\,\textmu m of nominal diameter at $\mathrm{NA}=0.2$), and it also led in nine of the ten seeds at equal solver time. The physics guidance added on top of the simulation feedback made no detectable difference to the span (mean differences of $+0.003$ and $-0.002$\,fs), and the advantage belongs to the LLM lanes as implemented, including their surrogate-ranked selection. An exploratory search seeded with every cell of the benchmark, in which cells designed by the agent set both ends of the library, extended the span to 3.46\,fs, close to a waveguide estimate for 600\,nm pillars. The nominal achromatic diameter at $\mathrm{NA}=0.2$ thus grows from 16.7\,\textmu m for the re-certified standard library to 20.5\,\textmu m (19.9\,\textmu m with the end-cell delays from full-wave FDTD). In full-lens simulations, the four lenses of the size ladder focus 54.1--56.5\% of the incident light on band average, with focal spreads of 2.01--2.12\,\textmu m, less than a fifth of the depth of focus. The value of the LLM lanes lay in reaching the ends of the design space within a fixed number of proposals, repeatedly across seeds. The campaign's short-delay cells, however, also follow from a conventional sweep once its range is extended. With prespecified hypotheses, paired seeds, four non-LLM baselines and explicit denominators, the benchmark offers a template for judging LLM agents in photonic design.

\section{Methods}
\label{sec:exp}

\subsection{Certification Ladder}
\label{sec:x-ladder}
The design campaign ran from day~0 to day~33, and the benchmark and the exploratory search followed on days 37--40, all days counted from the start of the campaign. Unit-cell responses were computed at normal incidence from the substrate by RCWA\cite{Moharam1981}, with the optical constants of atomic-layer-deposited TiO$_2$\cite{Zhukovsky2015} and of fused silica\cite{Malitson1965}. The campaign used the RCWA solver of Ansys Lumerical 2024~R1, whereas the benchmark and the exploratory search used our open RCWA solver. This solver, a Fourier modal method, reproduces the Lumerical RCWA spectra within 0.005--0.015\,fs in delay (Sec.~S2.2 in the Supplementary Material). In the screen, the certification ladder [Fig.~\ref{fig:cells}(e) and Table~S1 in the Supplementary Material] requires a cross-polarized conversion of at least 0.10 at 470, 570 and 670\,nm (bright) and convergence in the number $K$ of Fourier harmonics ($K$-converged). The $45^\circ$ rotation test allows a delay change of at most 0.25\,fs and a geometric-phase error of at most 0.35\,rad. The 21-wavelength spectrum requires a conversion of at least 0.10 at every wavelength, no phase step above $\pi/2$ and $\sigma_\varphi \le 0.10$\,rad. The certified delay $\tau$ is the linear coefficient of a least-squares fit of $\arg J_{\pm\mp}$ at the 21 wavelengths to a quadratic in $\Delta\omega$, and $\sigma_\varphi$ is the rms residual of that fit. A three-wavelength estimate, $\tau_3$, is used only in screening and in the rotation test. The gates, fixed from agent version V18 onward, do not bound the GDD, so a cell with any GDD can pass. Fabrication rules set minimum feature sizes, gaps and corner radii and an envelope for the rotated mask (Sec.~S2.3 in the Supplementary Material).

\subsection{Agent and Lanes}
\label{sec:x-agent}
In the campaign, an in-loop planner (gpt-5.4-mini) selected allowed simulation tasks from summaries of completed results. An in-loop inventor (gpt-6-sol) proposed new geometries in rounds of 20--25, first within parametric families (agent version V20) and then mostly as free-form outlines of up to three smoothed polygons (V21 and V22). Neither model could run code, alter the physics or relax a certification gate. In the final rounds (V22), the inventor drafted 60 candidates and kept 20 after seeing their predicted properties. Each random-mutation control has the same kind and target as its LLM proposal. A predictor-only candidate (V22) is the best of about 150 mutations of the 15 best certified cells as ranked by a $k$-nearest-neighbor surrogate trained on all 51{,}179 screens (held-out mean absolute error 0.09\,fs). A newly certified cell is useful if it extends the delay range of the certified pool before the free-form rounds by more than 0.02\,fs or is placed in the agent's scalar lens designs. It is also useful if its band-mean conversion exceeds by more than 0.01 that of every cell of that pool within $\pm0.10$\,fs of its delay [Eq.~(S13) in the Supplementary Material]. The prospective protocol of the campaign was not followed in full, so all campaign comparisons are exploratory (Sec.~S3.3 in the Supplementary Material). Proportions, with Wilson score 95\% intervals\cite{Wilson1927}, were compared with two-sided two-proportion $z$-tests and Fisher's exact tests, without adjustment for multiple comparisons. Matched pairs were compared with exact\cite{McNemar1947} and clustered\cite{Durkalski2003} McNemar tests over inventor rounds or control lineages (Sec.~S3.4 in the Supplementary Material).

\subsection{Benchmark Design and Statistics}
\label{sec:x-bench}
The protocol, code, starting pool and analysis of the benchmark were frozen together, with a SHA-256 manifest, before its confirmatory runs, but the protocol was not deposited in a public registry (Sec.~S6 in the Supplementary Material). The span of a library is the longest run of its certified delays in which neighboring delays differ by at most 0.25\,fs, the gap rule of the lens designs. The primary endpoint is the span of a seed's starting pool together with the certified cells of one lane-run after eight rounds of 20 proposals. Every lane-run sees only its own results and has the same fabrication and duplicate checks, the same certification and a budget of 160 proposals.

Certification uses the gates of the campaign with our open RCWA solver, and only candidates whose three-wavelength delay lies near the ends of the library ($\tau_3 \le 3.45$\,fs or $\tau_3 \ge 5.95$\,fs) receive the 21-wavelength test, by the same rule for every lane. The 1452-shape standard grid was re-certified once with this solver and ladder. Each seed $s = 1$--5 fixes a stratified random 80\% subsample of it as the starting pool of all lanes, so the comparison is paired within seeds. The LLM lane with physics guidance is the campaign's final agent in two-stage rounds. In each round, the inventor model, gpt-6-sol at high reasoning effort, states two to six testable physical hypotheses and drafts 60 cells. For each valid draft, it receives the prediction of the campaign's surrogate, retrained on the lane's own data, together with the three nearest simulated cells. It then chooses or revises 20 cells and updates a notebook of supported, refuted and open hypotheses. Its brief contains the lane's own data, summaries of simulated geometry--response trends and a physical description of the task, but none of the campaign's lessons. The LLM lane without physics guidance uses the same model, rounds, platform description, data tables and surrogate, but no physical description, trend summaries, hypotheses or notebook. Random mutation perturbs the five shortest- or five longest-delay cells, and the predictor-only lane simulates the best of 150 such mutations as ranked by the surrogate. Bayesian optimization selects batches by constrained expected improvement\cite{Jones1998} with the Kriging believer\cite{Ginsbourger2010}, or uses a trust region (TuRBO-1)\cite{Eriksson2019} in the free-form experiment, and CMA-ES\cite{Hansen2001} runs one instance per family and end. The free-form experiment, in which every lane draws one to three closed outlines of 3--8 points, was added as protocol revision~5 and frozen before its own runs while the parametric runs were in progress (Sec.~S6 in the Supplementary Material).

The unit of analysis is the lane-run, and in each seed the comparator is the best of the four non-LLM lanes, a strong requirement. H1 is tested with a one-sided exact paired sign-flip permutation test\cite{Good2005} on the five per-seed differences ($\alpha = 0.05$). Its smallest attainable $P$ is $1/32 = 0.031$, so H1 can be supported only if the LLM lane leads in every seed. H1-phys uses the same test in fixed sequence, that is, it is tested only if H1 is supported. This procedure keeps the probability of any false positive over the two tests at or below $\alpha$ without a correction for multiple testing, and each experiment is its own family of tests. Secondary endpoints are descriptive, with percentile bootstrap intervals over seeds (10{,}000 resamples). The span at equal solver time and gap rules of 0.15 and 0.35\,fs are sensitivity analyses. For the former, every lane-run is cut at the smallest nominal solver time of its seed, with each solve counted at the CPU time per solve that a self-test measured for its number of harmonics.

\subsection{Exploratory Search}
\label{sec:x-search}
After the benchmark, and outside it, the LLM lanes alone searched for a wider library. The starting pool merged every cell simulated in both experiments and their pilots, from all lanes and seeds but not from the campaign, deduplicated by shape, which gave 7183 cells, 3998 of them certified. Four lane-runs, with and without physics guidance for each of two new seeds, proposed free-form cells in the benchmark's two-stage rounds. The lane-runs with physics guidance started from the notebooks of the widest guided lane-runs of both experiments. The search ran 12 rounds, was extended to 20 while its ends still moved, and then continued in blocks of six rounds. It stopped after 32 rounds, when a block widened the span of the combined library by less than 0.005\,fs, and cells were certified as in the benchmark. The three shortest- and three longest-delay certified cells were then recomputed as periodic unit cells in Lumerical FDTD 2024~R1 with mesh overrides of 15\,nm laterally and 20\,nm vertically and of 10\,nm in all directions (Sec.~S2.5 in the Supplementary Material).

\subsection{Lens Design and Full-Wave Simulation}
\label{sec:x-lens}
The four lenses of the size ladder have the largest nominal diameters of their libraries [Eq.~\eqref{eq:dmax}], truncated to 0.1\,\textmu m. Lens~S uses the 181 cells of the re-certified standard grid, and Lenses~P and F add to it every certified cell of the parametric (1623 cells) or free-form (2194 cells) benchmark, from all lanes, seeds and pilots. Lens~W uses all 5975 certified cells, including the 1977 of the exploratory search. All four were assembled by one rule, fixed in the scalar model before the full-wave simulations. At each site of the 400\,nm lattice inside the aperture, the rule places the brightest cell within $\pm0.04$\,fs of the required delay, or the cell of nearest delay where no cell lies within this window. The placed cell is rotated to set the phase at 570\,nm (Sec.~S1.4 in the Supplementary Material).
Cells certified in the other circular channel are placed as their mirror images (Sec.~S4 in the Supplementary Material). Before the full-wave simulations, each design was evaluated in a scalar model that builds the aperture field from the certified responses of its cells in the locally periodic approximation\cite{Pestourie2018} and propagates it with the angular-spectrum method\cite{Goodman2017}. Full-lens simulations used Ansys Lumerical FDTD 2024~R1 with a mesh override of 15\,nm laterally and 20\,nm vertically over the pillars. Relative to the finest mesh tested, this mesh changes the group delay of ten test cells by at most 0.051\,fs (Fig.~S6 in the Supplementary Material). The converted field $(E_x \mp iE_y)/\sqrt{2}$ was propagated by the angular-spectrum method, and PSFs, Strehl ratios, MTFs and encircled energies were computed directly from the complex near fields, without phase fitting or cell averaging. The focal position at each wavelength is the intensity-weighted centroid of the main on-axis lobe above half its peak. The focal spread is its range over the band, and the common image plane is the band mean of the focal positions. The focusing efficiency is the power within a radius of 1.5 times the FWHM about the focus, divided by the incident or transmitted power (Sec.~S4 in the Supplementary Material). This FWHM is the diffraction-limited value in the scalar model and, in full wave, the simulated value in the plane of the on-axis intensity maximum.

\subsection{Optical-Design Analysis}
\label{sec:x-zemax}
The phase of the converted field at each wavelength was fitted with a six-term Binary~2 profile, $\varphi(\rho) = \sum_{i=1}^{6} A_i\,\rho^{2i}$ with $\rho = r/R$. The lenses were then analyzed in Zemax OpticStudio 2024~R1 through the ZOS-API at field angles of $0^\circ$, $10^\circ$ and $20^\circ$. Quick Focus (rms wavefront, centroid reference) gave the best focus at each wavelength, and the common image plane is their band mean (Sec.~S4.4 in the Supplementary Material).

\subsection{Use of Artificial Intelligence and Computing Resources}
\label{sec:x-ai}
Of the four artificial intelligence (AI) models used, OpenAI Codex (Codex desktop app running gpt-6-astra at medium or extra-high reasoning effort, used before the campaign and until day~18) was the coding and literature assistant for the atlas, the geometry grammar, agent versions V1--V14, the supervisor, the protocol and the literature component. Source summaries that it wrote for the atlas are labeled as AI-generated, and every reference cited here was verified against publisher or Crossref records. The in-loop planner was gpt-5.4-mini (OpenAI API), and the in-loop inventor was gpt-6-sol (OpenAI API, high reasoning effort, V20--V22), with 78 requests and a configured fallback to gpt-5.4 that was never used. The same model and settings drafted and chose the cells of both LLM lanes of the benchmark (320 requests) and of the exploratory search (257 requests), without a planner model or a fallback. Claude (Anthropic, claude-opus-5-5 in all logged sessions from day~19, with the model used on days 17 and 18 not recorded) was the development and campaign-management assistant from day~17 onward, working under the authors' direction. It implemented agent versions V15--V22.7 and the benchmark harness, built the FDTD and analysis pipelines, carried out the statistical analysis and verification and produced the figures (computing resources, model requests, tokens and costs in Sec.~S3.1 in the Supplementary Material).

Generative AI (gpt-5.4-mini, gpt-6-sol, OpenAI Codex and Claude) was part of the scientific method and the research infrastructure in the roles described above, including writing the plotting code that draws the figures from the simulation records. Claude (Anthropic, claude-opus-5-5) was used under the authors' direction to draft, restructure and edit the text of the manuscript and Supplementary Material, including figure captions, from the authors' results, analyses and decisions, to revise it in response to the authors' review comments and to check the reported numbers against the simulation records. The authors directed all work, reviewed and edited all text and take full responsibility for the content. No AI tool is an author, and no figure contains generative-AI imagery.

\subsection*{Disclosures}
The authors declare that there are no financial interests, commercial affiliations, or other potential conflicts of interest that could have influenced the objectivity of this research or the writing of this paper.

\subsection*{Code, Data, and Materials Availability}
Data underlying the results presented in this paper are available in Ref.~\citenum{Dataset1}. They comprise the meta-atom atlas and geometry catalog, all screening records, the proposals of the three lanes with their matched pairs, outcomes, lineage and stored surrogate predictions, including rejected and unevaluated candidates, the solver job records with their durations, the certified unit-cell library with its spectra, the final surrogate model and its validation report, the lens designs, the full-lens near fields, the frozen benchmark protocol with its manifests and amendment, every proposal, solve, model request and reply of the benchmark and the exploratory search, the starting pools, the unit-cell FDTD check and the inputs and outputs of every analysis. The agent, simulation, analysis and figure code, including our open RCWA solver and the conventional fin-pair search, together with the prospective protocol, its deviations and the recorded prompts, briefs and replies of the in-loop models with their model identifiers and settings, is available in Ref.~\citenum{Code1}. Commercial software (Ansys Lumerical and Ansys Zemax OpticStudio) is required to rerun some simulations and analyses and is not redistributed.

\subsection*{Acknowledgments}
This work was supported by the National Science Foundation (award 2414729). The use of generative AI in this work, including Claude (Anthropic) to draft, restructure and edit the text under the authors' direction, is described in Sec.~\ref{sec:x-ai}.

\bibliography{references}   %
\bibliographystyle{spiejour}   %

\begin{spacing}{1}
\end{spacing}

\end{spacing}
\clearpage
\includepdf[pages=-]{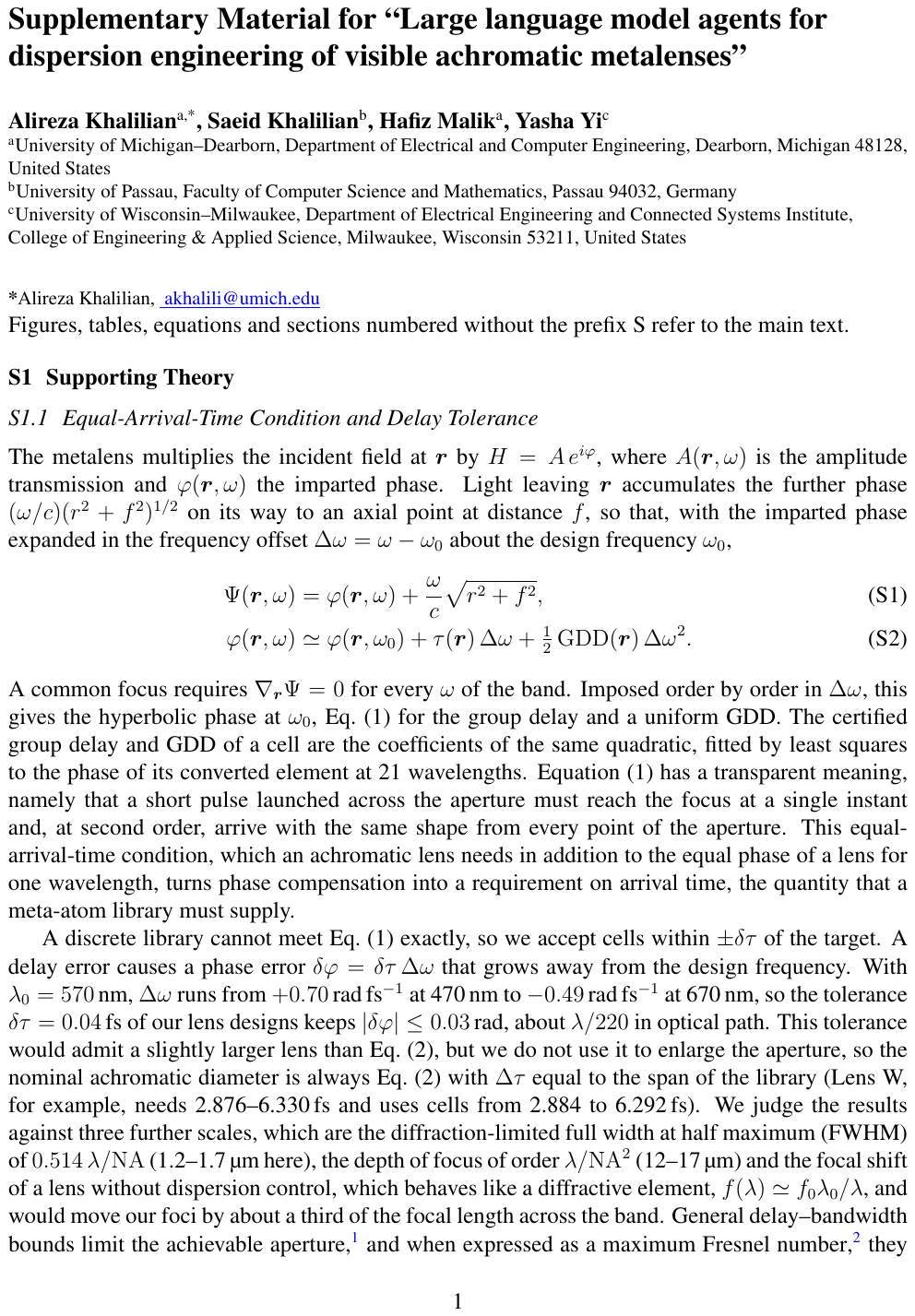}
\end{document}